# Distributive Subband Allocation, Power and Rate Control for Relay-Assisted OFDMA Cellular System with Imperfect System State Knowledge

Ying Cui, *Student Member*, Vincent K. N. Lau, *Senior Member* and Rui Wang, *Student Member, IEEE*

*Abstract*— In this paper, we consider distributive subband, power and rate allocation for a two-hop transmission in an orthogonal frequency-division multiple-access (OFDMA) cellular system with fixed relays which operate in decode-and-forward strategy. We take into account of system fairness by considering weighted sum goodput as our optimization objective. Based on the cluster-based architecture, we obtain a fast-converging distributive solution with only local imperfect CSIT by using decomposition of the optimization problem. To further reduce the signaling overhead and computational complexity, we propose a reduced feedback distributive solution, which can achieve asymptotically optimal performance for large number of users with arbitrarily small feedback overhead per user. We also derive asymptotic average system throughput for the relay-assisted OFDMA system so as to obtain useful design insights.

## I. INTRODUCTION

The relay-assisted OFDMA cellular system is a promising architecture for future wireless communication systems because it offers huge potential for enhanced system capacity, coverage as well as reliability [1], [2]. Full-duplex relay stations were discussed in the early literature. However, full-duplex transceiver is hard to implement in practical systems because the relay has to transmit and receive at that same time[1]. Since practical relays operate in a half-duplex manner, there is a duplexing penalty associated when using the relay to forward packets. As a result, it is not always advantageous to deliver packets via relay stations and it is very important to adapt the relay resource dynamically according to the global channel state information (GCSI)[2] in order to capture the advantage of the relay-assisted OFDMA architecture. However, perfect knowledge of GCSI is very difficult to obtain in both TDD (implicit CSI feedback) and FDD (explicit CSI feedback) systems due to the huge signaling overhead involved in delivering GCSI to the controller.

There are a lot of research interests focused on improving the system throughput by the optimal resource allocation of the relay-assisted OFDMA system. For example, in [3], [4], optimal subband allocation is considered for different scheduling schemes based on equal power allocation across all the subbands. In [5], heuristic separate subband and power allocation schemes are considered to improve the system capacity. To further improve the system throughput, joint subband and power allocation is proposed in [6], [7]. Due to the subband allocation constraint in the OFDMA system, the joint optimization problem is a NP-hard integer programming problem and continuous relaxation as well as graph theoretical approach are used to tackle the problem. However, as a simplification in both [6], [7], it is assumed that the source-relay, source-destination and relay-destination links use the same subband. They cannot effectively exploit the multiuser diversity, which is a very important component to system performance. While these works provide important initial investigations on the potential benefit of relay-assisted OFDMA systems, centralized resource allocation solutions and perfect knowledge of GCSI are required. In general, there are still several important remaining technical issues to be resolved in order to bridge the gap between theoretical gains and practical implementation considerations. They are elaborated below.

- **Distributive Implementation**: There are two potential issues associated with centralized implementations, namely the *complexity issue* and the *signaling loading issue*. For instance, the centralized joint optimization has a huge computational loading to the BS. Similarly, large signaling overhead is needed to collect the GCSI (BS-RS, BS-MS, RS-MS) from the RSs and MSs as well as to broadcast the scheduled results to the RSs and MSs. In [8], the authors proposed two semi-distributed sub-optimal subband allocation methods based on equal power allocation, which have offloaded certain computational load from the BS, but substantial signaling overhead is still needed to collect the GCSI from the RSs.
- **Imperfect Knowledge of GCSI**: While all the above works assume perfect GCSI knowledge at the transmitter, the CSIT measured at the transmitter side cannot be perfect due to either the CSIT estimation noise or the outdatedness of CSIT resulting from duplexing delay. When the CSIT is imperfect, there will be systematic packet errors (despite the use of strong channel coding) as long as the scheduled data rate exceeds the instantaneous channel mutual information. As a result, rate adaptation (in addition to power and subband allocations) must be considered in systems with imperfect CSIT in order to take into account of potential packet errors due to channel outage [9], [10], [11], [12].
- **Fairness Consideration**: Most of existing works only focus on sum-throughput maximization. However, fairness among users in the system is also an important consideration. In particular, it is important to study the potential fairness advantage of relay-assisted OFDMA systems as well.
- **Analytical Performance Results for Design Insights**: In all the existing works above, the system performance is obtained by simulations only. However, it is very difficult to obtain useful design insights (e.g. how the system performance scales with system parameters such as $M$, $K$ as well as path loss exponents of BS-RS and RS-MS links) without analytical performance results.

[1]In practice, there are cross-coupling between the transmit path and the receive path in any transceiver circuit. Hence, when the relay transmit (with high power), there will be some leakage into the receiver path, which will cause strong interference to the received signal (which is much weaker than the transmit signal).

[2]Global CSI refers to the CSI of the base station (BS) and relay (RS), CSI of the RS and mobile (MS) as well as the CSI of the BS and the MS.

In this paper, we shall attempt to shed some lights on the above issues. We consider a two-hop relay-assisted OFDMA system in a single cell with one base station (BS), $M$ relay stations (RS) and $K$ mobile stations (MS). In addition, we account for the penalty of packet errors due to imperfect CSIT by considering system goodput (b/s/Hz successfully delivered to the MS) as our performance measure instead of traditional sum-ergodic capacity (which only measures b/s/Hz sent by the BS or RS). We take into account of system fairness by considering weighted sum goodput as our optimization objective (which includes proportional fair scheduling (PFS) as a special case). Based on the cluster-based architecture, we obtain a fast-converging distributive solution with only local CSIT using careful decomposition of optimization problem. For the downlink implementation, to further reduce the signaling overhead and computational complexity, we propose a reduced feedback distributive solution and show that only an arbitrarily small feedback overhead per MS is needed to achieve asymptotically optimal performance for large $K$. Finally, we also derive asymptotic average system throughput for the relay-assisted OFDMA system so as to obtain useful design insights.

## II. SYSTEM MODEL

In this section, we shall describe the cluster-based system architecture, the imperfect GCSI model and the system utility.

### A. Cluster-based Architecture and Channel Model

Fig.1 illustrates the system model of the relay-assisted OFDMA system with one BS, $M$ fixed RSs and $K$ MSs. The coverage area is divided into $M+1$ clusters with cluster 0 served by the BS and cluster $m$ $(m \in \{1, M\})$ served by the $m$th RS. The $K$ MSs are assigned to one of the $M+1$ clusters according to their large scale path loss[3]. The number of MSs in cluster $m$ is $K_m$. MSs in cluster 0 will receive downlink packets or transmit uplink packets directly from or to the BS, and MSs in cluster $m$ $(m \in \{1, M\})$ will rely on the $m$th RS forwarding the downlink or uplink packets in the packet transmission process. Let $\mathcal{K}_m$ $(m \in \{1, M\})$ denote the set of MSs in cluster $m$, and $\mathcal{K}_0$ denote the set of MSs in cluster 0 and the RSs. For notation convenience, we assume index $k$ in cluster 0 denotes the $k$th MS if $k \in \{1, K_0\}$ and the $m$th RS if $k = K_0 + m$ $(m \in \{1, M\})$.

[3]The $k$th user is assigned to the cluster with the strongest received pilot strength.

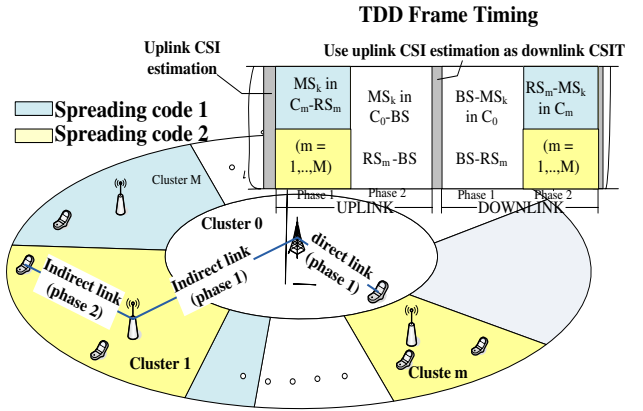

Fig. 1.   System Model

We consider frequency selective fading where there are $N$ independent multipaths. OFDMA is employed to convert the frequency selective fading channel into $T$ orthogonal subcarriers with $N$ independent subbands. The RSs operate in a half-duplex manner using decode-and-forward (DF) strategy. In order to facilitate relay-assisted packet transmission, a physical frame in the downlink is divided into two phases:
- In phase one, the BS transmits data to the MSs and the RSs belonging to cluster 0.
- In phase two, the BS stops transmitting while the RSs (which have successfully decoded data from the BS in phase one) will forward the data to the MSs belonging to their own clusters.

Similarly, the two phases of a physical frame in the uplink are
- In phase one, the MSs belonging to the relay clusters will transmit data to their own RSs.
- In phase two, the MSs and the RSs (which have successfully decoded data from their MSs in phase one) belonging to cluster 0 will transmit the data to the BS.

Two orthogonal frequency spreading codes are assigned to adjacent RS clusters to mitigate potential mutual interference between them as illustrated in Fig.1[4]. Since the path loss between non-adjacent RS clusters are usually quite large, there is practically negligible mutual interference between non-adjacent RS clusters. As a result, all the RS clusters can operate simultaneously on the entire frequency band with negligible interference[5].

[4]To achieve the orthogonal signal separation in the frequency domain, signals from/to different RSs must be synchronized in the OFDM symbol boundary. This requires timing synchronization within the unused cyclic prefix in the OFDMA system and is implementable in practice. As a result, it is a common assumption in OFDMA systems such as WiMAX.

In two phases ($m = 0$ for the phase one of the downlink (phase two of the uplink) and $m \in \{1, M\}$ for the phase two of the downlink (phase one of the uplink)), the received symbol $Y_{m,k,n}$ carrying user $k$' information in cluster $m$ in the $n$th subband is given by

$$Y_{m,k,n} = \sqrt{p_{m,k,n} l_{m,k}} H_{m,k,n} X_{m,k,n} + Z_{m,k,n} \; m \in \{0, M\}$$

where $X_{m,k,n}$ is the transmitted symbol from (to) BS ($m = 0$) or the $m$th RS $(m \in \{1, M\})$ to (from) user $k$ $(k \in \mathcal{K}_m)$ in cluster $m$ in the $n$th subband, $p_{m,k,n}$ is the transmit SNR, $l_{m,k}$ is the path loss, and $Z_{m,k,n} \sim \mathcal{CN}(0, 1)$ is the noise in the $n$th subband. We consider Rayleigh fading and hence, $H_{m,k,n} \sim \mathcal{CN}(0, 1)$.

### B. Global Channel State Information Model

In this paper, we consider a relay-assisted cellular system with large coverage, and hence, both small scale fading (due to multipath) and large scale fading (due to path loss) are considered. We consider the block fading channel where the small scale fading coefficient is quasi-static within each frame and may be different among frames. We consider the resource allocation for the low-mobility users. Since the time scale for mobility is much larger than that for small scale fading, the path loss remains constant for a large number of frames and can be estimated with high accuracy [6]. The CSIT of small scale fading can be obtained from either explicit feedback (FDD systems) or implicit feedback (TDD systems) using reciprocity between uplink and downlink[7]. We consider TDD systems, and assume perfect CSIR and imperfect CSIT due to estimation noise on the uplink pilots or duplexing delay as illustrated in Fig. 1. The estimated CSIT in phase one and phase two can be modeled as

$$\hat{H}_{m,k,n} = H_{m,k,n} + \triangle H_{m,k,n} \quad m \in \{0, M\} \tag{1}$$

where $H_{m,k,n}$ is the actual CSI, and $\triangle H_{m,k,n} \sim \mathcal{CN}(0, \sigma_e^2)$ is the CSIT error. We denote the set of local imperfect CSIT of cluster $m$ as $\hat{\boldsymbol{H}}_m = \{\hat{H}_{m,k,n} | k \in \mathcal{K}_m, \forall n\}$, and the global imperfect CSIT as $\hat{\boldsymbol{H}} = \bigcup_{m=0}^{M} \hat{\boldsymbol{H}}_m$.

### C. System Policy

In this paper, we consider joint subband, power and rate allocation[8]. The system policies are defined below.

*1) Subband Allocation Policy $\mathcal{S}$:* Let $s_{m,k,n}^B \in \{0, 1\}$ be the subband allocation indicator at the BS for MS $k$ in cluster $m$ ($k \in \{1, K_0\}$ when $m = 0$, $k \in \mathcal{K}_m$ when $m \in \{1, M\}$). Let $s_{m,k,n}^R \in \{0, 1\}$ be the subband allocation indicator at the $m$th $(m \in \{1, M\})$ RS for user $k$ $(k \in \mathcal{K}_m)$ in cluster $m$. The subband allocation policy is

$$\mathcal{S} = \Big\{ s_{m,k,n}^B, s_{m,k,n}^R \in \{0,1\} | \forall n, \sum_{m=1}^{M} \sum_{k \in \mathcal{K}_m} s_{m,k,n}^B$$
$$+ \sum_{k=1}^{K_0} s_{0,k,n}^B = 1, \sum_{k \in \mathcal{K}_m} s_{m,k,n}^R = 1 \, \forall m \in \{1, M\} \Big\}$$

*2) Power Allocation Policy $\mathcal{P}$:* Let $p_{m,k,n}$ be the scheduled transmit SNR at BS ($m = 0$) and the $m$th RS $(m \in \{1, M\})$ respectively to user $k$ $(k \in \mathcal{K}_m)$ in cluster $m$ in the $n$th subband. Let $P_m$ be the total transmit SNR at BS ($m = 0$) and the $m$th RS $(m \in \{1, M\})$. Let $P_{m,k}$ be the total transmit SNR at the user $k$ $(k \in \mathcal{K}_m)$. The power allocation policy in downlink and uplink systems ($\forall m \in \{0, M\}$) are

$$\mathcal{P}^{DL} = \Big\{ p_{m,k,n} \geq 0 \, | \, \sum_{n=1}^{N} \sum_{k \in \mathcal{K}_m} p_{m,k,n} \leq P_m, \forall m \Big\}$$

$$\mathcal{P}^{UL} = \Big\{ p_{m,k,n} \geq 0 \, | \, \sum_{n=1}^{N} p_{m,k,n} \leq P_{m,k}, \forall m, k \in \mathcal{K}_m \Big\}$$

*3) Rate Allocation Policy $\mathcal{R}$:* Let $r_{m,k,n}$ be the scheduled data rate at BS ($m = 0$) and the $m$th RS $(m \in \{1, M\})$ respectively to user $k$ $(k \in \mathcal{K}_m)$ in cluster $m$ in the $n$th subband. The rate allocation policy is

$$\mathcal{R} = \Big\{ r_{m,k,n} \geq 0 \, | \, \forall n, m \in \{0, M\}, k \in \mathcal{K}_m \Big\}$$

### D. Maximum Achievable Date Rate and System Goodput

In this part, we shall introduce the system utility based on the notations and system policies defined in the previous part[9].

Given perfect CSIR, the instantaneous mutual information (bit/s/Hz) between the $m$th transmitter (receiver) and the $k$th receiver (transmitter) ($k \in \mathcal{K}_m$) in the $n$th subband is given by

$$C_{m,k,n} = c_m \log_2(1 + p_{m,k,n} l_{m,k} |H_{m,k,n}|^2) \tag{2}$$

where $c_0 = 0.5$ and $c_m = 0.25$ $(m \in \{1, M\})$[10].

Due to imperfect CSIT, the transmitter does not have knowledge on the instantaneous mutual information in (2) and hence, the scheduled data rate at the BS and the RS might be larger than the corresponding mutual information, leading to packet outage. To take the potential outage into consideration and to guarantee the scheduling fairness, we consider the average weighted system goodput (average weighted b/s/Hz successfully delivered to the MSs) as our performance measure. The average weighted total system goodput is given by $\bar{U}_{wgp}(\mathcal{S},\mathcal{P},\mathcal{R}) = \mathbf{E}_{\hat{\mathbf{H}}}[U_{wgp}(\mathbf{S},\mathbf{P},\mathbf{R},\hat{\mathbf{H}})]$, where $U_{wgp}(\mathbf{S},\mathbf{P},\mathbf{R},\hat{\mathbf{H}})$ is the conditional average total weighted goodput for given $\hat{\mathbf{H}}$, and is given by

$$U_{wgp}(\mathbf{S},\mathbf{P},\mathbf{R},\hat{\mathbf{H}}) = \frac{1}{N}\Bigg(\sum_{m=1}^{M}\sum_{k\in\mathcal{K}_m}\alpha_k^m \min$$

$$\Bigg\{\sum_{n=1}^{N} s_{m,k,n}^B r_{0,K_0+m,n}(1-\Pr[C_{0,K_0+m,n}<r_{0,K_0+m,n}|\hat{\mathbf{H}}]),$$

$$\sum_{n=1}^{N} s_{m,k,n}^R r_{m,k,n}(1-\Pr[C_{m,k,n}<r_{m,k,n}|\hat{\mathbf{H}}])\Bigg\}$$

$$+\sum_{k=1}^{K_0}\alpha_k^0\sum_{n=1}^{N} s_{0,k,n}^B r_{0,k,n}(1-\Pr[C_{0,k,n}<r_{0,k,n}|\hat{\mathbf{H}}])\Bigg) \quad (3)$$

where $\alpha_k^m$ denotes the weight[11] of the MS $k$ in Cluster $m$, and $\Pr[C_{m,k,n}<r_{m,k,n}|\hat{\mathbf{H}}]$ ($\forall n, m\in\{0,M\}, k\in\mathcal{K}_m$) is the conditional packet error probability of one-hop link for given global $\hat{\mathbf{H}}$.

---

[5]For tractable analysis, we assumed all RSs are separated either spatially or on the code domain. Yet, the effect of mutual interference is taken into consideration in the performance simulation.

[6]For example, in 802.16e system, frame duration is 5ms. Users with pedestrian mobility 5 km/hr will have coherence time at least over 20 frames.

[7]In practical systems, like IEEE 802.16e, a mechanism named "channel sounding" is proposed to enable the BS to determine the BS-to-MS channel response under the assumption of TDD reciprocity.

[8]For both the uplink and downlink transmission, the resource allocation is performed at BS and RSs.

[9]Note that $p_{0,k,n}$, $r_{0,k,n}$ ($k = K_0 + m$) are the power and rate control variables for the $m$th the RS ($m = 1, \cdots, M$). $s_{m,k,n}^B$ is the subband allocation variable for the direct link BS-MS$_k$ with $m = 0$, $k \in \{1, \cdots, K_0\}$. $s_{m,k,n}^B, s_{m,k,n}^R$ are the subband allocation variables for indirect link BS-RS$_m$-MS$_k$ with $m \in \{1, M\}$, $k \in \mathcal{K}_m$.).

[10]$\frac{1}{2}$ is due to duplexing penalty. $\frac{1}{4}$ is due to duplexing penalty and the spreading code with code rate $\frac{1}{2}$ used in RS clusters. For simplicity, we assume that the same amount of time resources is used for each phase. However our design can be directly applied on the system with unequal phase duration by changing the duplexing penalty ratio between phase one and phase two.

### III. SUBBAND, POWER AND RATE SCHEDULING PROBLEM FORMULATION

In this section, we shall formulate the relay-assisted scheduling problem as an optimization problem maximizing the average total weighted goodput $\bar{U}_{wgp}(\mathcal{S},\mathcal{P},\mathcal{R})$ subject to the target outage probability $\epsilon$. Note that optimization on $\bar{U}_{wgp}(\mathcal{S},\mathcal{P},\mathcal{R})$ w.r.t. policies (set of actions for all CSIT realizations) is equivalent to optimization on $U_{wgp}(\mathbf{S},\mathbf{P},\mathbf{R},\hat{\mathbf{H}})$ w.r.t. the actions $\mathbf{S}$, $\mathbf{P}$, $\mathbf{R}$ ($\mathbf{S}=\mathcal{S}(\hat{\mathbf{H}})$, $\mathbf{P}=\mathcal{P}(\hat{\mathbf{H}})$, $\mathbf{R}=\mathcal{R}(\hat{\mathbf{H}})$) for each given CSIT realization. Hence, the optimization problem is given by

*Problem 1:* (*Subband, Power and Rate Optimization Problem*)

$$\max_{\mathbf{S},\mathbf{P},\mathbf{R}} U_{wgp}(\mathbf{S},\mathbf{P},\mathbf{R},\hat{\mathbf{H}})$$

s.t. $\Pr[C_{m,k,n}<r_{m,k,n}|\hat{\mathbf{H}}]=\epsilon \,\forall n,m\in\{0,M\},k\in\mathcal{K}_m$ (4)

$$s_{m,k,n}^B, s_{m,k,n}^R \in\{0,1\}\,\forall n,m\in\{0,M\},k\in\mathcal{K}_m \quad (5)$$

$$\sum_{m=1}^{M}\sum_{k\in\mathcal{K}_m} s_{m,k,n}^B + \sum_{k=1}^{K_0} s_{0,k,n}^B = 1\,\forall n \quad (6)$$

$$\sum_{k\in\mathcal{K}_m} s_{m,k,n}^R = 1\,\forall n, m\in\{1,M\} \quad (7)$$

$$p_{m,k,n}\geq 0\,\forall n,m\in\{0,M\},k\in\mathcal{K}_m \quad (8)$$

$$\sum_{n=1}^{N}\sum_{k\in\mathcal{K}_m} p_{m,k,n}\leq P_m, \quad \forall m\in\{0,M\}\,\text{(for DL)} \quad (9\text{a})$$

$$\sum_{n=1}^{N} p_{m,k,n}\leq P_{m,k}, \quad \forall m\in\{0,M\}\,\text{(for UL)} \quad (9\text{b})$$

---
[11]Proportional fairness (PF) is a special case of the weighted sum goodput system utility.

### IV. DUAL PROBLEM AND DISTRIBUTIVE SOLUTION

Note that Problem 1 is a mixed integer real optimization and hence, is not convex. Brute-force optimization is NP-hard with exponential complexity in term of the number of subbands. In this section, we shall apply continuous relaxation technique [6], [3], [13], [14], to obtain asymptotically optimal solution as well as discuss the distributive implementation.

#### A. Continuous Relaxation of the Integer Programming Problem

To perform continuous relaxation, we allow time sharing between users for each subband by relaxing subband allocation indicator to rational value between 0 and 1, which describes the fraction of time a particular subband is occupied by a particular user. Mathematically, to apply the continuous relaxation, the constraint (5) is replaced by

$$s_{m,k,n}^B, s_{m,k,n}^R \geq 0 \quad \forall n,m\in\{0,M\},k\in\mathcal{K}_m \quad (10)$$

After equivalent transformation and continuous relaxation, Problem 1 is approximated by the following convex optimization problem.

*Problem 2 (Relaxed Optimization Problem):*

$$\max_{\mathbf{S},\mathbf{P},\mathbf{t}}\sum_{m=1}^{M}\sum_{k\in\mathcal{K}_m}\alpha_k^m t_k^m + \sum_{k=1}^{K_0}\alpha_k^0\sum_{n=1}^{N} s_{0,k,n}^B \tilde{r}_{0,k,n}^B$$

s.t. $t_k^m \leq \sum_{n=1}^{N} s_{m,k,n}^B \tilde{r}_{m,k,n}^B \quad m\in\{1,M\},k\in\mathcal{K}_m \quad (11)$

$t_k^m \leq \sum_{n=1}^{N} s_{m,k,n}^R \tilde{r}_{m,k,n}^R \quad m\in\{1,M\},k\in\mathcal{K}_m \quad (12)$

as well as constraints in $(10),(6),(7),(8),(9\text{a})$ for DL or $(9\text{b})$ for UL

where[12]

$$\tilde{r}_{m,k,n}^B = \frac{1}{2}\log_2\Big(1+\frac{p_{m,k',n}\hat{g}_{m,k',n}}{s_{m,k,n}^B}\Big), m\in\{0,M\}$$

$(m=0, k'=k\in\{1,K_0\}; m>0, k'=K_0+m, k\in\mathcal{K}_m)$

$$\tilde{r}_{m,k,n}^R = \frac{1}{4}\log_2\Big(1+\frac{p_{m,k,n}\hat{g}_{m,k,n}}{s_{m,k,n}^R}\Big), m\in\{1,M\},k\in\mathcal{K}_m$$

$$\hat{g}_{m,k,n} = l_{m,k}\frac{1}{2}\sigma_e^2 F^{-1}_{|\hat{H}_{m,k,n}|^2/\frac{1}{2}\sigma_e^2}(\epsilon), m\in\{0,M\},k\in\mathcal{K}_m$$

(13)

Note that Problem 2 is a convex problem. For the objective function, the first part $\sum_{m=1}^{M}\sum_{k\in\mathcal{K}_m}\alpha_k^m t_k^m$ is linear in $t_k^m$, and the second part $\sum_{k=1}^{K_0}\alpha_k^0\sum_{n=1}^{N} s_{0,k,n}^B \tilde{r}_{0,k,n}^B$ is a positive linear combination of functions of the type $f(x,y)=x\log(1+y/x), y\geq 0, x\geq 0$. According to Lemma 1 of [14], the second part is concave in $(s_{0,k,n}^B, p_{0,k,n})$. Therefore, the objective function is concave. Similarly, after moving the R.H.S. of constraint (11) and (12) to the L.H.S., it can be proved that the inequality constraint functions are convex. In addition, the left equality constraint functions are linear. Hence, Problem 2 is a convex optimization problem.

---
[12]$F^{-1}_{|\hat{H}_{m,k,n}|^2/\frac{1}{2}\sigma_e^2}(\cdot)$ denotes the inverse cdf of non-central chi-square random variable with 2 degrees of freedom and non-centrality parameter $|\hat{H}_{m,k,n}|^2/\frac{1}{2}\sigma_e^2$. (13) can be derived from the conditional PER constraint (4), the expression of instantaneous mutual information (2) and the non-central chi-square distribution of $|H_{m,k,n}|^2$ given $\hat{H}_{m,k,n}$, which is omitted due to page limit.

The continuous relaxation in Problem 2 does not necessarily yield a solution where all the subband allocation indicators are 0 or 1. However, they are 0 or 1 with high probability due to the property of marginal benefit of extra bandwidth defined in Appendix B. If the integer solution is required, we can round the fractional values to 0 or 1. In fact, under some mild condition ($\frac{T}{N} \to \infty$) [13], the solution of Problem 2 will be identical to that of Problem 1 when we do subcarrier allocation in both problems.

*Remark 1:* Let $\frac{T}{N}$ be the number of subcarriers in each independent subband. We assume the channel gains of all the subcarriers in one subband are highly correlated. Suppose a user is assigned $n (0 \leq n \leq \frac{T}{N})$ subcarriers in a particular subband during one transmission. We can interpret $n/\frac{T}{N}$ as the frequency sharing factor in this subband, which takes discrete value from 0 to 1. When $\frac{T}{N} \to \infty$, the frequency sharing factor can take any rational between 0 and 1. It can be achieved by the further subcarrier allocation within the subband.

### B. Partial Dual Decomposition and Distributed Solution

The subband, power and bit rate allocations for relay-assisted OFDMA system is a complicated optimization problem. Existing approach [3], [6], [7] considered centralized solutions but these solutions have huge computational loading at the BS. Furthermore, these solutions require knowledge of GCSI which induces heavy signaling overhead in the system. In this section, we shall derive a low-complexity distributive solution from Problem 2 using decomposition techniques.

Using convex optimization techniques (details in Appendix A), the dual function of Problem 2 can be simplified as follows

$$\max_{\boldsymbol{S},\boldsymbol{P}} \sum_{k=1}^{K_0} \alpha_k^0 \sum_{n=1}^{N} s_{0,k,n}^B \tilde{r}_{0,k,n}^B + \sum_{m=1}^{M} \sum_{k \in \mathcal{K}_m} \lambda_k^m \sum_{n=1}^{N} s_{m,k,n}^B \tilde{r}_{m,k,n}^B$$
$$+ \sum_{m=1}^{M} \sum_{k \in \mathcal{K}_m} (\alpha_k^m - \lambda_k^m) \sum_{n=1}^{N} s_{m,k,n}^R \tilde{r}_{m,k,n}^R \quad (14)$$

s.t. $(10), (6), (7), (8), (9a)$ for DL or $(9b)$ for UL

In what follows, we apply dual decomposition approach [15], [16]. The dual function can be decomposed into two subproblems:

*Subproblem 1:* (Resource Allocations at BS in Phase One of Downlink (Phase two of Uplink)) Given $M$ vectors of Lagrangian multipliers $\boldsymbol{\lambda^m}$, optimize the subband and power allocations at BS such that the weighted goodput is maximized subject to corresponding subband constraints at the BS, and power constraints at the BS for downlink (at the MSs in cluster 0 and RSs for uplink).

$$g_{BS}(\boldsymbol{\lambda^1}, \cdots, \boldsymbol{\lambda^M})$$
$$= \begin{cases} \max_{\boldsymbol{S},\boldsymbol{P}} \sum_{m=1}^{M} \sum_{k \in \mathcal{K}_m} \lambda_k^m \sum_{n=1}^{N} s_{m,k,n}^B \tilde{r}_{m,k,n}^B \\ \quad + \sum_{k=1}^{K_0} \alpha_k^0 \sum_{n=1}^{N} s_{0,k,n}^B \tilde{r}_{0,k,n}^B \\ \text{s.t. } s_{m,k,n}^B \geq 0, (6), (8), (9a) \text{ or } (9b) \ (m = 0) \end{cases}$$

*Subproblem 2:* (Resource Allocations at the $m$th RS in Phase Two of Downlink (Phase one of Uplink)) Given the vector of Lagrangian multipliers $\boldsymbol{\lambda^m}$, optimize subband and power allocations at the $m$th RS such that the weighted goodput is maximized subject to corresponding subband constraints at the $m$th RS, and power constraints at the $m$th RS for downlink (at the MSs in cluster $m$ for uplink).

$$g_{RS}^m(\boldsymbol{\lambda^m})$$
$$= \begin{cases} \max_{\boldsymbol{S},\boldsymbol{P}} \sum_{k \in \mathcal{K}_m} (\alpha_k^m - \lambda_k^m) \sum_{n=1}^{N} s_{m,k,n}^R \tilde{r}_{m,k,n}^R \\ \text{s.t. } s_{m,k,n}^B \geq 0, (7), p_{m,k,n} \geq 0, (9a) \text{ or } (9b) \ (m > 0) \end{cases}$$

There are $M$ subproblems of this kind, each one corresponding to the resource allocations at one RS.

Subproblem 1 and Subproblem 2 are similar. Please refer to the Appendix B for the optimal solution. With the dual function mentioned above, the dual problem is summarized below:

*Problem 3 (Dual Problem):* Find the optimal dual variables which maximize the dual function

$$\min_{\boldsymbol{\lambda}} g_{BS}(\boldsymbol{\lambda^1}, \cdots, \boldsymbol{\lambda^M}) + \sum_{m=1}^{M} g_{RS}^m(\boldsymbol{\lambda^m})$$
$$\text{s.t. } \boldsymbol{0} \preccurlyeq \boldsymbol{\lambda^m} \preccurlyeq \boldsymbol{\alpha^m}, m \in \{1, M\}$$

where $\boldsymbol{\alpha^m} = (\alpha_k^m)_{K_m \times 1} (m \in \{1, M\})$.

We use the subgradient method [17] to update each dual variable as follows

$$\lambda_k^m(i+1) \quad (15)$$
$$= \left[ \lambda_k^m(i) - \delta_k^m(i) (\sum_{n=1}^{N} s_{m,k,n}^B \tilde{r}_{m,k,n}^B - \sum_{n=1}^{N} s_{m,k,n}^R \tilde{r}_{m,k,n}^R) \right]_{\mathcal{X}_k^m}$$

where $\delta_k^m(i)$ is a positive step size and $[\cdot]_{\mathcal{X}_k^m}$ denotes the projection onto the feasible set $\mathcal{X}_k^m = \{\lambda_k^m | 0 \leq \lambda_k^m \leq \alpha_k^m\}$ (i.e. let $\lambda_k^m(i+1) = 0 \forall [\cdot] < 0$, $\lambda_k^m(i+1) = \alpha_k^m \forall [\cdot] > \alpha_k^m$ and keep $\lambda_k^m(i+1) = [\cdot]$ if $[\cdot] \in \mathcal{X}_k^m$). We study the convex Problem 2 by solving its dual problem with decomposition technique and subgradient method. Since the Problem 2 is convex and strictly feasible[14], Slater's condition holds. Therefore, duality gap is zero and hence, the primal variables $\boldsymbol{S}(\boldsymbol{\lambda^1}(i), \cdots, \boldsymbol{\lambda^M}(i))$ and $\boldsymbol{P}(\boldsymbol{\lambda^1}(i), \cdots, \boldsymbol{\lambda^M}(i))$ will converge to the primal optimal variables $\boldsymbol{S^*}, \boldsymbol{P^*}$. Since the dual problem is always convex, the convergence of the proposed scheme is guaranteed [17]. For any initial value in the feasible set, the dual variable $\lambda_k^m(i)$ will converge to the dual optimal $\lambda_k^{m*}$ as $i \to \infty$.

The distributive architecture of the derived solution is illustrated in Fig.2. Intuitively, for user $k$ in cluster $m$ ($m \in$

---

[13] The condition is quite mild and can be satisfied in most practical systems. For example, we have $N = 2048$ and $T \approx 6$ in WiMAX (802.16e) systems.

[14] For example, we can easily find $s_{m,k,n}^B, s_{m,k,n}^R > 0$, which satisfy (6) and (7), and $p_{m,k,n} > 0$, which satisfy (9a) for downlink or (9b) for uplink. By choosing $t_k^m = \min\{\sum_{n=1}^{N} s_{m,k,n}^B \tilde{r}_{m,k,n}^B, \sum_{n=1}^{N} s_{m,k,n}^R \tilde{r}_{m,k,n}^R\} - \epsilon (\forall \epsilon > 0)$, we can obtain a strictly feasible solution to the convex problem 2.

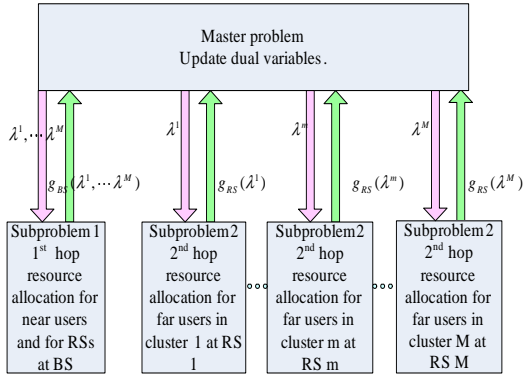

Fig. 2. Decomposition Structure

$\{1, \cdots, M\}$), the dual variable $\lambda_k^m(i)$ can be interpreted as the equivalent weight in phase one for downlink (phase two for uplink), while $\alpha_k^m - \lambda_k^m$ as the equivalent weight in phase two for downlink (phase one for uplink). Based on the weights assigned by the master problem, Subproblem 1 can be solved at BS with its local imperfect CSIT ($\hat{\mathbf{H}}_0$), and Subproblem 2 for the $m$th RS can be solved at the $m$th RS with its local imperfect CSIT ($\hat{\mathbf{H}}_m$). Then the dual problem updates the dual variables at BS to reduce the difference between the scheduled data rate for a particular user in phase one and phase two at each iteration.

The distributive implementation offloads great computations from the BS to $M$ RSs. In addition, it saves large signaling overhead for collecting CSI of RS-MS links and broadcasting the scheduled results to RSs for resource allocations of RS-MS links. These advantages are highly desirable for practical implementation.

## V. REDUCED FEEDBACK DISTRIBUTIVE SOLUTION FOR DOWNLINK SYSTEMS

Compared with the centralized solution, the complexity and signaling overload in the above distributive solution are greatly reduced. However, in downlink systems, the overhead due to the feedback of CSI from MSs to their cluster controllers (the BS and RSs) and the signaling between the BS and RSs still grows linearly with the number of users $K$[15]. Selective multiuser diversity (SMUD) has been proposed in [18], [19] for cellular system to reduce feedback overhead by imposing a local threshold-based screening. We shall extend the threshold-based mechanism to reduce the overall system feedback overhead of the distributive solution for relay-assisted cellular network with fairness consideration. The reduced feedback distributive solution is outlined below[16].

---
[15]In uplink systems, since resource allocation is performed at the BS and RSs, which are the receivers, we do not have the issues of CSI feedback overhead. After performing distributive resource allocation, the BS and active RSs inform the corresponding scheduled MSs with the scheduled subband and transmit SNR for transmission.

[16]In practice (e.g. HSDPA, WiMAX, 802.16m, LTE), there is always a pre-allocated dedicated UL signaling channel associated with high speed downlink packet data access and DL resource allocation. So our feedback information from MS->RS and RS->BS can be transmitted over these available dedicated control channels.

*Algorithm 1 (Distributive Algorithm with Reduced Feedback):*

1) The controller of cluster m ($m \in \{0, M\}$) finds out MS $i$ with the maximum weight $\alpha_i^m$ among MSs in Cluster $m$, and broadcasts the feedback threshold $\gamma_{th}^m = \alpha_i^m l_i^m \tilde{\gamma}^m$ to the MSs in its cluster.
2) Each MS in Cluster $m$ compares and feedbacks its CSI $h_{k,n}^m$ in the $n$th subband iff $\alpha_k^m l_k^m |h_{k,n}^m|^2 \geq \gamma_{th}^m$. Then the local reduced user set and corresponding imperfect CSIT[17] is available at each cluster manager.
3) BS decides and broadcasts the initial multipliers $\{\boldsymbol{\lambda^m}(0) | m \in \{1, M\}\}$ in phase one. The initial multipliers $\{\boldsymbol{\alpha^m} - \boldsymbol{\lambda^m}(0) | m \in \{1, M\}\}$ in phase two are available at RSs for subproblem 2.
4) In the $i$th iteration, BS solves Subproblem 1 and each RS solves its own Subproblem 2. Each RS reports the scheduled data rate of users in its cluster to BS.
5) BS updates the multipliers $\{\boldsymbol{\lambda^m}(i)\}$ to $\{\boldsymbol{\lambda^m}(i+1)\}$ in phase one according to (15), and broadcasts $\{\boldsymbol{\lambda^m}(i+1)\}$.
6) If the difference of the scheduled data rate in two phases for each user in RS clusters is less than a threshold, or the number of iterations has already reached a predetermined value, then terminate the algorithm. Otherwise, jump to step 4).

In the above algorithm, the communication overhead between BS and each RS grows linearly only with the total number of users in $M+1$ reduced user sets. Hence, the feedback load of this algorithm is much smaller than the directly distributive implementation with full feedback. Although this algorithm is in general suboptimal due to the existence of feedback outage, it can be proved in the following lemma that under some conditions, its performance will converge to that of the directly distributive implementation with all MSs transmitting feedbacks. Due to the symmetric situation in each subband, we only consider one subband and ignore the index $n$ for simplicity.

*Lemma 1:* (*Feedback Outage, Feedback Load and Asymptotic Performance*) Assume that the weight of each user is not smaller than 1. Without loss of generality, assume user $i$ has the maximum weight $\alpha_i^m$ in Cluster $m$. Define $T_k^m = \frac{\alpha_i^m l_i^m}{\alpha_k^m l_k^m}$.

1) Given the threshold $\gamma_{th}^m = \alpha_i^m l_i^m \tilde{\gamma}^m$, the outage probability $P_0^m$ (the probability that no one feedbacks) and the feedback load $F^m$ (average number of feedbacks per user) are given by $P_0^m = \prod_{k=1}^{K_m}(1 - \exp(-T_k^m \tilde{\gamma}^m))$ and $F^m = \frac{1}{K_m} \sum_{k=1}^{K_m}(1 - \exp(-T_k^m \tilde{\gamma}^m))$ respectively.
2) Let the upper bound of $P_0^m$ be $\overline{P_0^m}(K_m)$. If $\overline{P_0^m}(K_m)$ satisfies $\overline{P_0^m}(K_m) \to 0$ and $\overline{P_0^m}(K_m)^{\frac{1}{K_m}} \to 1$ as $K_m \to \infty$, choose $\tilde{\gamma}^m(\overline{P_0^m}(K_m)) = \frac{1}{\max T_k^m} \log \frac{1}{1 - \overline{P_0^m}(K_m)^{\frac{1}{K_m}}}$, so that $P_0^m \to 0$ and $F^m \to 0$ as $K_m \to \infty$.

There is a tradeoff between feedback outage probability and feedback load by adjusting $\tilde{\gamma}^m$. However, for sufficiently large

---
[17]The reduced user set is made up of users who feedback at least in one subband. The channel coefficient of any user in this set will be treated as 0 in the subbands without feedback from this user.

$K_m$, Algorithm 1 can achieve asymptotically optimal weighted average goodput at asymptotically zero feedback cost per user by choosing $\tilde{\gamma}^m(\overline{P_0^m}(K_m)) = \frac{1}{\max T_k^m}\log\frac{1}{1-\overline{P_0^m}(K_m)^{\frac{1}{K_m}}}$.

*Proof:* Please refer to Appendix C. ∎

## VI. ASYMPTOTIC PERFORMANCE ANALYSIS UNDER PFS FOR DOWNLINK SYSTEMS

In this section, we shall focus on studying how the system performance of the scheduling algorithm derived in the preceding section for downlink systems grow with various important system parameters such as the number of RSs and the number of MSs. Specifically, we consider proportional fair scheduling (PFS) [20] performance in the limit situation $t_c \to \infty$ [21], [22] for sufficiently large $K$ in the analysis. To obtain insights on the performance gains, we impose a set of simplifying assumptions. We assume each RS contains $K$ MSs with $K_0 = 0$. Furthermore, we assume line-of-sight link (with high gain antenna) between the RSs and the BS and hence, the throughput is limited by the second hop. Under above assumptions, we can substantially simplify the system model without affecting the asymptotic performance and hence, we can derive closed form results which would otherwise be impossible in general regimes.

Let $D$ be the radius of a cell. Path loss takes on the form $PL(d) = d^{-\alpha}$, where $\alpha$ is the path loss exponent. In the relay-assisted system, assume there are $M$ RSs and $K$ users uniformly distributed in each RS cluster. Assume equal power allocation to each subband at RSs[18]. Accordingly, in the system without RS, we assume $MK$ users in the same cell edge region as the relay clusters, which is equivalent to the total number of users in the relay-assisted system, and BS allocates equal power to each subband. The asymptotic performance under PFS of the two systems is summarized in the following lemma.

*Lemma 2:* (*Asymptotic Performance for the Systems with RSs and without RSs*) For the system with $M$ RSs and $K$ MSs in each RS-cluster, the average asymptotic throughput for large $K$ is $E[T] = \frac{M}{4}\Big(\log_2(\frac{P_m}{N}\ln K) - \alpha(\log_2 D + \log_2 t - \frac{1}{2\ln 2})\Big)$. For the system with $MK$ MSs and no RSs, the average asymptotic throughput for large $K$ is $E[T^{(b)}] = \log_2(\frac{P_0}{N}\ln MK) - \alpha(\log_2 D + \frac{(1-2t)^2}{1-(1-2t)^2}\log_2\frac{1}{1-2t} - \frac{1}{2\ln 2})$, where $t = \frac{\sin\frac{\pi}{M}}{1+\sin\frac{\pi}{M}}$.

*Proof:* Please refer to Appendix D. ∎

Define the gain of relay-assisted design over the non-relay design as $g = E[T] - E[T^{(b)}] = g_{fsr} + g_{pl}$, where $g_{fsr} = \frac{M}{4}(\log_2(\frac{P_m}{N}\ln K) - \alpha\log_2 D) - (\log_2(\frac{P_0}{N}\ln(MK)) - \alpha\log_2 D)$ is the throughput gain for frequency and spacial reuse and $g_{pl} = \alpha\Big(\frac{(1-2t)^2}{4t-4t^2}\log_2\frac{1}{1-2t} - \frac{M}{4}\log_2 t + (\frac{M}{4} - 1)\frac{1}{2\ln 2}\Big)$ is the throughput gain for reducing energy reduction due to path loss.

*Remark 2:* Since $g_{fsr} = \mathcal{O}(M\ln\ln K) - \mathcal{O}(\ln\ln(MK))$, the throughput of relay-assisted system grows much faster than the system without RS, which is due to the spatial reuse of relay-assisted architecture. It can be shown that $g_{pl} > 0$ and increases with $M$. This demonstrates the benefits of relay-assisted architecture on reducing energy reduction.

---
[18]Since the number of MSs are large, equal power allocation is asymptotically optimal due to multi-user diversity.

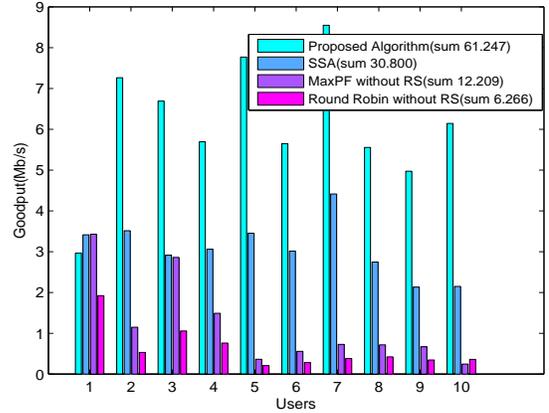

Fig. 3. Average goodput allocation for 10 users in a single cell scenario at BS transmit power 36 dBm, RS transmit power 36 dBm at $N = 16$.

## VII. SIMULATION RESULTS AND DISCUSSION

In this part, we shall compare our distributive subband, power and rate allocation for relay-assisted OFDMA system with several baseline references. Baseline 1 refers to the weighted total goodput maximization version of Separate and Sequential Allocation (SSA), which is a semi-distributed scheme proposed in [8]. Baseline 2 refers to the maximum total weighted goodput scheduling without RSs. In Baseline 3, we consider Round Robin Scheduling with water-filling power allocation across the subbands. We apply PFS algorithm to keep track of the average goodput of each user, and consider its inverse as the weight for the three maximum total weighted goodput scheduling design. We use Jain's index as the fairness measure, which ranges from $1/K$ (worst case) to 1 (best case). In the following simulation results, the average sum goodput is obtained from the optimal scheduled goodput of each user. The cell radius is 5000m. 6 RSs are evenly located on the circle with radius 3000m. We set up our simulation scenarios according to the practical settings in IEEE802.16m systems [23]. The carrier frequency is 2GHz. BS/RS height is 32m and MS height is 1.5m. The operating bandwidth is 10MHz with 2048 subcarriers and 8, 16 or 24 independent subbands. The path loss model of BS-MS and RS-MS is $128.1 + 37.6\log_{10}(R)$ dB, and the path loss model BS-RS is $128.1 + 28.8\log_{10}(R)$ dB ($R$ in km). The receive antenna gains of MS is 0 dB, and the directional receive antenna is used at RS with antenna gain 20 dB. The lognormal shadowing standard deviation is 8 dB, and the CSIT error variance $\sigma_e^2$ is 0.01.

### A. System Performance of the Distributive Algorithm

Fig.3 illustrates the average scheduled goodput allocation of the 10 users in a single cell scenario, the positions of which are generated according to uniform distribution. It can be observed that our distributive scheduling algorithm can achieve much higher average goodput and fairness compared with three baselines.

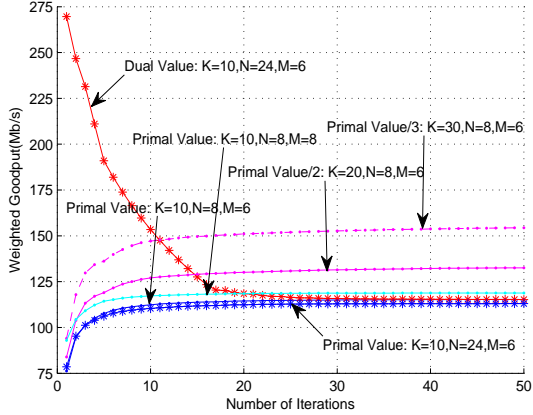

Fig. 4. Average total weighted goodput in a single cell scenario for primal problem and dual problem of 10 users (N=24, M=6, K=10) and primal problem of 4 other cases (N=8, M=8, K=20,30 respectively) versus the number of iterations at BS transmit power 36 dBm, RS transmit power 36 dBm. (Due to limited space in the graph, we have omitted the dual values of the other 4 cases.)

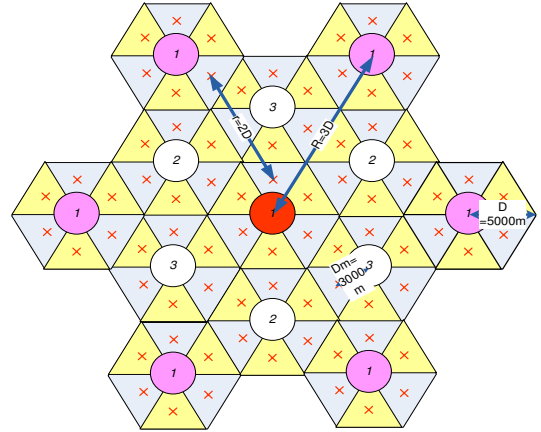

Fig. 5. Multi-Cell Scenario with frequency reuse factor 3 (The number in the cell center shows the frequency band used by that cell.). Two orthogonal frequency spreading codes are assigned to adjacent RSs to achieve the orthogonal signal separation in the frequency domain (blue and yellow parts represent two frequency spreading codes adopted in RS clusters.). The cell radius $D$ is 5000m. 6 RSs are evenly located on the circle with radius $D_m$ =3000m. The distance between the BSs of the nearest co-channel cells is $R = 3D$. The distance between the RSs of the nearest relay clusters of the nearest co-channel cells is $r = 2D$.

Fig.4 illustrates the convergence performance of our distributive algorithm in a single cell scenario. We plot the average best primal value curve and dual value curve within certain number of iterations for the 10 users case ($N = 24, M = 6, K = 10$) and the best primal value curve for four other cases with $N = 8, M = 8, K = 20, 30^{19}$, respectively. It can be seen that our distributive algorithm converges quite fast. The performance at the 5th iteration is about 95% of the performance at the 50th iteration. Thus, good performance can be achieved with low overhead.

### B. System Performance versus Transmit Power

Fig.6 illustrates the average sum goodput performance of the 10 users in a multi-cell scenario with frequency reuse factor 3 as illustrated in Fig. 5 versus the transmit power at BS and RS. It can be observed that our proposed scheduling design has significant gain, especially in the lower SNR region. This is mainly because RS reduces the path loss greatly and our proposed algorithm utilizes the limited power more efficiently.

Fig.7 illustrates the fairness performance of the 10 users in a multi-cell scenario versus the transmit power. It can be seen that the designs with RSs keep much better fairness, especially in the lower SNR region. The main reason is that the differences of path loss of users in the cell are greatly reduced when the RS is half-way in between the BS and the MS. Compared with Baseline 1, our proposed design has similar fairness performance but a much better goodput performance as illustrated in Fig.6.

[19]The convergence performance for the full feedback distributive algorithm with 30 MSs, is actually the same as the convergence performance for the reduced feedback distributive algorithm with 150 MSs under only 5% goodput performance penalty.

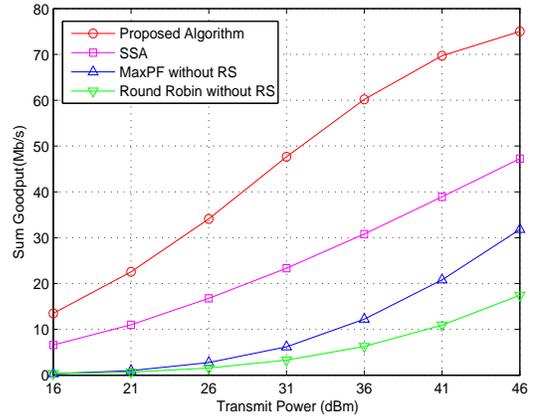

Fig. 6. Average sum goodput of 10 users versus BS/RS transmit power in a multi-cell scenario with frequency reuse factor 3 as illustrated in Fig. 5 at $N = 16$.

### C. System Performance versus the Number of Users K

Fig.8 illustrates the average sum goodput performance versus the number of users per RS cluster in a single cell scenario. The transmission in phase two of our design directly benefit from the increase in the number of cell-edge users. When BS and RSs have the same transmit power 36 dBm, the bottleneck of the performance is in phase one. Hence, the first curve dose not increase much with the number of users in RS clusters. When the transmit power at RSs is smaller than that at BS, i.e. 31 dBm, 26 dBm, the bottleneck shifts to phase two. Thus, the overall performance increases greatly with the number of users.

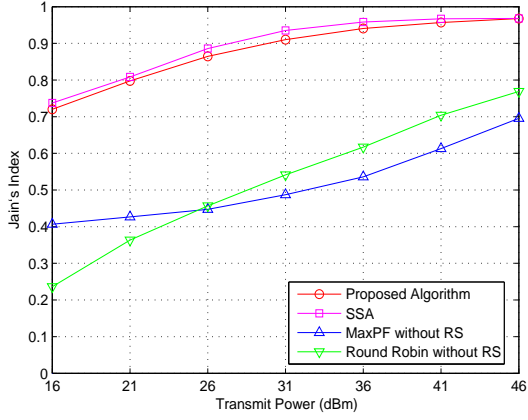

Fig. 7. Jain's fairness index of 10 users versus BS/RS transmit power in a multi-cell scenario with frequency reuse factor 3 as illustrated in Fig. 5 at $N = 16$.

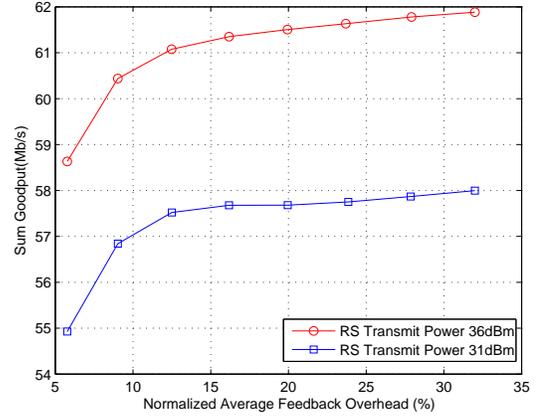

Fig. 9. Average sum goodput of 60 users per cell in a single cell scenario versus the average feedback load at BS transmit power 36dBm, RS transmit power 36dBm, 31dBm, and $N = 8$.

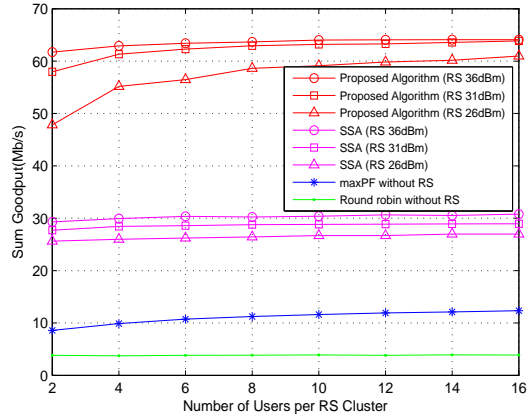

Fig. 8. Average sum goodput versus the number of users per RS cluster (6 RSs) in a single cell scenario at BS transmit power 36dBm, RS transmit power 36dBm, 31dBm, 26dBm, and $N = 8$.

### D. Tradeoff of Performance and Average Feedback Load in the Distributive Reduced Feedback Algorithm

Fig.9 illustrates the average sum goodput performance for 60 users per cell in a single cell scenario versus the average feedback load by implementing distributive reduced feedback algorithm. We use different feedback thresholds to generate different average sum goodput performance with different average feedback load. Our distributive reduced feedback algorithm can achieve good performance (e.g. 95 % of the distributive full feedback algorithm in Fig.8) with low feedback load (e.g. 20 %) in real systems.

### VIII. SUMMARY

In this paper, we propose a cluster-based distributive subband, power and rate allocation for a two-hop transmission in a relay-assisted OFDMA cellular system. We take into account of potential packet errors due to imperfect CSIT and system fairness by considering weighted sum goodput as our optimization objective. Based on the cluster-based architecture, we obtain a fast-converging distributive solution with only local imperfect CSIT by using careful decomposition of optimization problem. Our solution could be applied to both UL and DL allocations. To further reduce the signaling overhead and computational complexity in downlink systems, we propose a reduced feedback distributive algorithm, which can achieve asymptotically optimal performance for large number of users with arbitrarily small feedback outage and feedback load. We also derive asymptotic average system goodput for the relay-assisted OFDMA system so as to obtain useful design insights.

### APPENDIX A: DERIVATION OF DUAL FUNCTION

By relaxing the global coupling constraints (11) and (12), we have the following dual function

$$\max_{\boldsymbol{S},\boldsymbol{P},\boldsymbol{t}} \sum_{m=1}^{M} \sum_{k \in \mathcal{K}_m} \alpha_k^m t_k^m + \sum_{k=1}^{K_0} \alpha_k^0 \sum_{n=1}^{N} s_{0,k,n}^B \tilde{r}_{0,k,n}^B$$
$$+ \sum_{m=1}^{M} \sum_{k \in \mathcal{K}_m} \lambda_k^m (\sum_{n=1}^{N} s_{m,k,n}^B \tilde{r}_{m,k,n}^B - t_k^m)$$
$$+ \sum_{m=1}^{M} \sum_{k \in \mathcal{K}_m} \nu_k^m (\sum_{n=1}^{N} s_{m,k,n}^R \tilde{r}_{m,k,n}^R - t_k^m)$$
s.t.(10), (6), (7), (8), (9a)for DL or (9b)for UL

where $\boldsymbol{\lambda^m} \triangleq (\lambda_k^m)_{K_m \times 1} \succcurlyeq \boldsymbol{0}$, $\boldsymbol{\nu^m} \triangleq (\nu_k^m)_{K_m \times 1} \succcurlyeq \boldsymbol{0}$ ($m \in \{1, M\}$ are the vectors of Lagrangian multipliers. Firstly, we optimize over $t$. The part of dual function with respect to $t$ is given by

$$g_0 = \max_{\boldsymbol{t}} \sum_{m=1}^{M} \sum_{k \in \mathcal{K}_m} \left(\alpha_k^m - (\lambda_k^m + \nu_k^m)\right) t_k^m$$
$$= \begin{cases} 0, & \lambda_k^m + \nu_k^m = \alpha_k^m \\ \infty, & \text{otherwise} \end{cases}$$

To make sure that the dual function is bounded above, we have $\lambda_k^m + \nu_k^m = \alpha_k^m$. Hence, we can simplify the dual function.

## APPENDIX B: OPTIMAL SOLUTION TO THE RELAXED PROBLEM

For downlink systems, the subproblem 1 and 2 share the similar form[20] as follows

$$\max_{S,P} \sum_{k=1}^{K} \alpha_k \sum_{n=1}^{N} s_{k,n} \log_2\left(1 + \frac{g_{k,n} p_{k,n}}{s_{k,n}}\right)$$

$$\text{s.t.} s_{k,n} \geq 0, \sum_{k=1}^{K} s_{k,n} = 1 \forall n, p_{k,n} \geq 0, \sum_{n=1}^{N}\sum_{k=1}^{K} p_{k,n} \leq P$$

The Lagrangian function is given by $L = \sum_{k=1}^{K} \alpha_k \sum_{n=1}^{N} s_{k,n} \log_2\left(1 + \frac{g_{k,n} p_{k,n}}{s_{k,n}}\right) + \mu\left(P - \sum_{n=1}^{N}\sum_{k=1}^{K} p_{k,n}\right) + \sum_{n=1}^{N} v_n\left(1 - \sum_{k=1}^{K} s_{k,n}\right).$

$$\frac{\partial L}{\partial p_{k,n}} = 0 \Rightarrow p_{k,n} = s_{k,n}\left(\frac{\alpha_k}{\mu} - \frac{1}{g_{k,n}}\right)^+ \quad (16)$$

$$\frac{\partial L}{\partial s_{k,n}} = 0 \Rightarrow X_{k,n} \triangleq \alpha_k \log_2\left(1 + \frac{g_{k,n} p_{k,n}}{s_{k,n}}\right)$$
$$- \alpha_k \frac{g_{k,n} p_{k,n}}{s_{k,n} + g_{k,n} p_{k,n}} = v_n \quad (17)$$

where $X_{k,n}$ can be interpreted as marginal benefit of extra bandwidth [14]. By substituting (16) into (17), we have $X_{k,n} = \alpha_k \log\left(1 + g_{k,n}\left(\frac{\alpha_k}{\mu} - \frac{1}{g_{k,n}}\right)^+\right) - \mu\left(\frac{\alpha_k}{\mu} - \frac{1}{g_{k,n}}\right)^+$. $\mu$ satisfies

$$\sum_{k=1}^{K}\sum_{n=1}^{N} s_{k,n}\left(\frac{\alpha_k}{\mu} - \frac{1}{g_{k,n}}\right)^+ = P \quad (18)$$

$\mu$ can be obtained by the subgradient method. For a particular $\mu$, if there is a unique $k^* = \arg\max\{X_{k,n}\}$ for some $n$, time-sharing will not happen in this subband.

$$s_{k,n} = \begin{cases} 1, & X_{k,n} = \max_k\{X_{k,n}\} > 0 \\ 0, & \text{otherwise} \end{cases}$$

Since for each given $\mu$, $X_{k,n}$ is a function of the CSI $g_{k,n}$ of each user and they are independent random variable. As a result, there is probability 0 for $X_{k,n} = X_{k',n}$ with $k \neq k'$. Hence, there is probability 1 that the solution of the relaxed problem corresponds to that of the original problem[21].

For uplink systems, we use individual power constraint $\sum_{n=1}^{N} p_{k,n} \leq P_k$ instead. Similarly, we have $p_{k,n} = s_{k,n}\left(\frac{a_k}{\mu_k} - \frac{1}{g_{k,n}}\right)^+$, where $\mu_k$ satisfies $\sum_{n=1}^{N} s_{k,n}\left(\frac{a_k}{\mu_k} - \frac{1}{g_{k,n}}\right)^+ = P_k$ and $s_{k,n}$ can be obtained with the same method as the downlink.

## APPENDIX C: PROOF OF LEMMA 1

For notation convenience, we only consider MSs in one cluster and omit cluster index $m$ in the following proof. Let $g_k$ ($= l_k|h_k|^2$) be the channel gain of user $k$. Define $x_k \triangleq \alpha_k g_k$. Given $\alpha_i > \alpha_j \geq 1 \forall j \neq i$, we first show that if $\exists j$ such that $x_i > x_j$, then subband will not be allocated to user $j$ according to Appendix B.

$$x_i > x_j, \forall j \neq i \Rightarrow \begin{cases} \frac{1}{\mu} - \frac{1}{x_i} > \frac{1}{\mu} - \frac{1}{x_j} > 0 \\ \frac{1}{\mu} - \frac{1}{x_i} > 0 \geq \frac{1}{\mu} - \frac{1}{x_j} \\ 0 \geq \frac{1}{\mu} - \frac{1}{x_i} > \frac{1}{\mu} - \frac{1}{x_j} \end{cases}$$

In case 2 and 3, $p_j = 0$. It is obvious that user $j$ can not transmit in this subband. We only need to consider case 1. $\forall k$, if $\frac{1}{\mu} - \frac{1}{x_k} > 0$, $X_k(\alpha_k, x_k) = \alpha_k \log\left(\frac{x_k}{\mu}\right) - \alpha_k\left(1 - \frac{x_k}{\mu}\right)$. Now, we shall show that $X_k(\alpha_k, x_k)$ increase with $\alpha_k$ and $x_k$ separately. Let $t_k = x_k/\mu$, $\frac{\partial X_k}{\partial \alpha_k} = \log t_k - \left(1 - \frac{1}{t_k}\right) \triangleq f(t_k)$ and $\frac{\partial X_k}{\partial x_k} = \frac{\alpha_k x_k - \mu}{x_k^2}$. Then $\frac{1}{\mu} - \frac{1}{x_k} > 0 \Rightarrow t_k > 1 \Rightarrow f'(t_k) = \frac{1}{t_k} - \frac{1}{t_k^2} > 0 \Rightarrow f(t_k > 1) > f(t_k = 1) = 0 \Rightarrow \frac{\partial X_k}{\partial \alpha_k} > 0$. In addition, $\alpha_k \geq 1, x_k > \mu \Rightarrow \frac{\alpha_k x_k - \mu}{x_k^2} > \frac{x_k - \mu}{x_k^2} > 0 \Rightarrow \frac{\partial X_k}{\partial x_k} > 0$. Thus, $\alpha_i > \alpha_j, x_i > x_j > 0 \Rightarrow X_i(\alpha_i, x_i) > X_j(\alpha_j, x_j)$. According to Appendix C, this subband will not be allocated to user $j$.

Given the threshold $\gamma_{th} = \alpha_i l_i \tilde{\gamma}$, user $k$ will feedback its channel quality to the cluster manager iff $\alpha_k g_k \geq \alpha_i l_i \tilde{\gamma} \Rightarrow \gamma_k = |h_k|^2 \geq T_k \tilde{\gamma}$, where $T_k = \frac{\alpha_i l_i}{\alpha_k l_k}$. R.V. $\xi_k$ denotes the feedback action of user $k$ in one subband. $\xi_k = 1$, if user k feedback its channel quality; $\xi_k = 0$ otherwise. For Rayleigh fading assumed in the context, $\Pr[\xi_k = 1] = Pr[\gamma_k \geq T_k \tilde{\gamma}] = e^{-T_k \tilde{\gamma}}$, $Pr[\xi_k = 0] = Pr[\gamma_k \leq T_k \tilde{\gamma}] = 1 - e^{-T_k \tilde{\gamma}}$.

Intuitively, we can consider $l_i \tilde{\gamma}$ as the overall channel gain of a potential user who shares the same weight $\alpha_i$ as user $i$ and serve as a threshold user. The scheduling is done among the feedback user set and the threshold user. If $\tilde{\gamma} \leq |h_i|^2$, the scheduling is still optimal. If $\tilde{\gamma} > |h_i|^2$ and $\sum_{k=1}^{K} \xi_k > 0$, the scheduling is optimal except for the threshold user is scheduled at last, which happens with low possibility. If $\sum_{k=1}^{K} \xi_k = 0$, we declare a scheduling outage and suffer from certain performance loss. The outage probability is $P_0 = \Pr[\sum_{k=1}^{K} \xi_k = 0] = \prod_{k=1}^{K}(1 - \exp(-T_k\tilde{\gamma}))$. The average feedback load is $F = \frac{1}{K} E[\sum_{k=1}^{K} \xi_k] = \frac{1}{K}\sum_{k=1}^{K}(1 - e(-T_k\tilde{\gamma}))$. We have $\exp(-\max\{T_k\}\tilde{\gamma}) \leq \exp(-T_k\tilde{\gamma}) \leq \exp(-\min\{T_k\}\tilde{\gamma}) \Rightarrow P_0 \leq (1 - \exp(-\max\{T_k\}\tilde{\gamma}))^K, F \leq \exp(-\min\{T_k\}\tilde{\gamma})$. $(1 - \exp(-\max\{T_k\}\tilde{\gamma}))^K = \overline{P_0^m}(K) \Rightarrow \tilde{\gamma}(\overline{P_0^m}(K)) = \frac{1}{\max\{T_k\}} \log \frac{1}{1-\overline{P_0^m}(K)^{\frac{1}{K}}} \Rightarrow \exp(-\min\{T_k\}\tilde{\gamma}) = \exp\left(-\frac{\min\{T_k\}}{\max\{T_k\}} \log \frac{1}{1-\overline{P_0^m}(K)^{\frac{1}{K}}}\right) = \left(1 - \overline{P_0^m}(K)^{\frac{1}{K}}\right)^{-\frac{\min\{T_k\}}{\max\{T_k\}}}$. Since $\overline{P_0^m}(K) \to 0$ as $K \to \infty$, we have $P_0 \to 0$. Because $\overline{P_0^m}(K)^{\frac{1}{K}} \to 1$ as $K \to \infty$, we have $F \to 0$. Hence, $\alpha_i l_i \tilde{\gamma}(\overline{P_0^m}(K))$ is asymptotic optimal threshold. For example, $\overline{P_0^m}(K) = \frac{1}{K}$ satisfies all the conditions.

## APPENDIX D: PROOF OF LEMMA 2

By applying the similar approach as in [21], throughput (bit/s/Hz) for MS $k \in \mathcal{K}_m$ in relay-assisted system and in the

---

[20]For subproblem 1, different $(m, k)$ pairs can be treated as different $k$ in this form. Subproblem 2 for a given m is equivalent to this form by omitting index $m$.

[21]If $\exists k \neq k'$ such that $X_{k,n} = X_{k',n}$ is maximum, $s_{k,n}$ can be obtained by solving a set of equations $\sum_{k=1}^{K} s_{k,n} = 1$ and $X_{k,n} = X_{k',n}$ for the sharing subbands and (18).

system without RS are:

$$T_{m,k} = \frac{1}{4K} \int_0^\infty \log_2\left(1 + \frac{P_m}{N} l_{m,k} x dF_{\max,K}(x)\right)$$
$$\doteq \frac{1}{4K} \log_2\left(\frac{P_m}{N} l_{m,k} \ln K\right) \quad \text{for large } K$$
$$T_{m,k}^{(b)} = \frac{1}{MK} \int_0^\infty \log_2\left(1 + \frac{P_0}{N} l_{m,k}^{(b)} x dF_{\max,MK}(x)\right)$$
$$\doteq \frac{1}{MK} \log_2\left(\frac{P_0}{N} l_{m,k}^{(b)} \ln(MK)\right) \quad \text{for large } K$$

where $F_{\max,K}(x)$ and $F_{\max,MK}(x)$ are cdf of $\max\{|H_{m,k,n}|^2, \forall k\}$ and $\max\{|H_{m,k,n}|^2, \forall m, k\}$ separately, $l_{m,k}$ and $l_{m,k}^{(b)}$ are path loss from the RS $m$ and BS for the MS $k \in \mathcal{K}_m$ separately.

Next, we consider the average throughput over distance. To get closed form solution, we use closed disc with radius $D_M$ to approximate the relay clusters. The the relation between $D$ and $D_M$ is given by $(D - D_M)\sin\frac{\pi}{M} = D_M \Rightarrow D_M = \frac{\sin\frac{\pi}{M}}{1+\sin\frac{\pi}{M}} D = tD$. Assume MSs are uniformly distributed in a relay cluster and the system without RS. Therefore, we get the average total throughput in relay-assisted system and the system without RS for large $K$ in Lemma 2. In addition, it can be easily shown that $\frac{dg_{pl}}{dM} > 0$.

## References

[1] WINNER- Wireless World Initiative New Radio. [Online]. Available: http://www.ist-winner.org/.
[2] O. Oyman, J. N. Laneman, and S. Sandhu, "Multihop relaying for broadband wireless mesh networks: from theory to practice," *IEEE Commun. Mag.*, vol. 45, no. 11, pp. 116 – 122, Nov. 2007.
[3] W. Nam, W. Chang, S. Y. Chung, and Y. H. Lee, "Transmit optimization for relay-based cellular OFDMA systems," in *IEEE Int. Conf. on Commun. (ICC)*, Glasgow, Scotland, June 2007, pp. 5714 – 5719.
[4] O. Oyman, "Opportunistic scheduling and spectrum reuse in relay-based cellular OFDMA networks," in *IEEE Global Telecommunications Conference (GLOBECOM)*, Washington, DC.,USA, Nov. 2007, pp. 3699–3703.
[5] L. Huang, L. Wang, and E. Schulz, "Resource allocation for OFDMA based relay enhanced cellular networks," in *IEEE Veh. Tech. Conf. (VTC)*, Baltimore, USA, Apr. 2007, pp. 3160 – 3164.
[6] T. C. Y. Ng and W. Yu, "Joint optimization of relay strategies and resource allocations in cooperative cellular networks," *IEEE J. Select. Areas Commun.*, vol. 25, no. 2, pp. 328 – 339, Feb. 2007.
[7] G. Li and H. Liu, "Resource allocation for OFDMA relay networks," *IEEE J. Select. Areas Commun.*, vol. 24, no. 11, Nov. 2006.
[8] M. K. Kim and H. S. Lee, "Radio resource management for a two-hop OFDMA relay system in downlink," *IEEE Symposium on Computers and Communications*, pp. 25 – 31.
[9] V. K. Lau, W. K. Ng, and D.S.W. Hui, "Asymptotic tradeoff between cross-layer goodput gain and outage diversity in OFDMA systems with slow fading and delayed CSIT," *IEEE Trans. Wireless Commun.*, pp. 2732–2739, July 2008.
[10] F. Brah, L. Vandendorpe, and J. Louveaux, "OFDMA constrained resource allocation with imperfect channel knowledge," in *IEEE Veh. Tech. Conf. (VTC)*, Benelux, Nov. 2007, pp. 1–5.
[11] ——, "Constrained resource allocation in OFDMA downlink systems with partial CSIT," in *IEEE Int. Conf. on Commun. (ICC)*, Beijing, China, May 2008, pp. 4144 – 4148.
[12] I. C. Wong and B. L. Evans, "Optimal OFDMA subcarrier, rate, and power allocation for ergodic rates maximization with imperfect channel knowledge," *IEEE Trans. Acoust., Speech, Signal Processing*.
[13] L. M. C. Hoo, B. Halder, J. Tellado, and J. M. Cioffi, "Multiuser transmit optimization for multicarrier broadcast channels," *IEEE Trans. Commun.*, vol. 52, no. 6, June 2004.
[14] W. Yu and J. M. Cioffi, "FDMA capacity of gaussian multiple-access channels with ISI," *IEEE Trans. Commun.*, vol. 50, no. 1, Jan. 2002.
[15] D. P. Palomar and M. Chiang, "A tutorial on decomposition methods for network utillity maximization," *IEEE J. Select. Areas Commun.*, vol. 24, no. 8, pp. 1439 – 1451, Aug. 2006.
[16] ——, "Alternative decompositions and distributed algorithms for network utility maximization," in *IEEE Global Telecommunications Conference (GLOBECOM)*, vol. 5, St. Louis, Missouri, Nov. 2005, pp. 2563 – 2568.
[17] S. Boyd, L. Xiao, and A. Mutapcic, *Lecture notes of EE392o: Subgradient methods*. Standford Univ., 2003.
[18] D. Gesbert and M. S. Alouini, "How much feedback is multi-user diversity really worth," in *IEEE Int. Conf. on Commun. (ICC)*, vol. 1, Paris, France, 2004, pp. 234 – 238.
[19] V. Hassel, M. S. Alouini, and D. Gesbert, "Rate-optimal multiuser scheduling with reduced feedback load and analysis of delay effects," *EURASIP Journal on Wireless Communications and Networking*, pp. 1 – 7, 2006.
[20] P. Viswanath and D. Tse, "Opportunistic beamforming using dumb antennas," *IEEE Trans. Inform. Theory*, vol. 48, no. 6, pp. 1277 – 1294, June 2002.
[21] G. Caire, R. Muller, and R. Knopp, "Hard fairness versus proportional fairness in wireless communications: the single-cell case," *IEEE Trans. Inform. Theory*, vol. 53, no. 4, pp. 1366 – 1385, Apr. 2007.
[22] D. Park and D. Gesbert, "How much feedback is multi-user diversity really worth," in *International Symposium on Information Theory (ISIT)*, Toronto, Canada, 2008, pp. 2036 – 2040.
[23] IEEE 802.16m evaluation methodology document. IEEE 802.16m-08/004r5. [Online]. Available: http://www.ieee802.org/16/tgm/

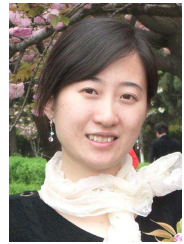

**Ying Cui** received B.Eng degree (first class honor) in Electronic and Information Engineering, Xian Jiaotong University, Xi'an, China in 2007. She is currently a Ph.D candidate in the Department of Electronic and Computer Engineering, the Hong Kong University of Science and Technology (HKUST). Her current research interests include cooperative and cognitive communications, delay-sensitive cross-layer scheduling as well as stochastic approximation and Markov Decision Process.

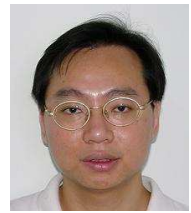

**Vincent K. N. Lau** obtained B.Eng (Distinction 1st Hons) from the University of Hong Kong in 1992 and Ph.D. from Cambridge University in 1997. He was with PCCW as system engineer from 1992-1995 and Bell Labs - Lucent Technologies as member of technical staff from 1997-2003. He then joined the Department of ECE, HKUST as Associate Professor. His current research interests include the robust and delay-sensitive cross-layer scheduling, cooperative and cognitive communications as well as stochastic approximation and Markov Decision Process.

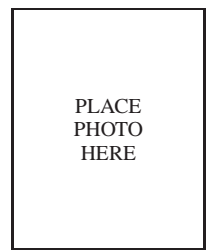

PLACE PHOTO HERE

**Rui Wang** received B.Eng degree (first class honor) in Computer Science from the University of Science and Technology of China in 2004 and Ph.D degree in the Department of ECE from HKUST in 2008. He is currently a post-doctoral researcher in HKUST. His current research interests include cross-layer optimization, wireless ad-hoc network, and cognitive radio.